\documentclass[conference]{IEEEtran}
\IEEEoverridecommandlockouts
\usepackage[utf8]{inputenc}
\usepackage[T1]{fontenc}
\usepackage{cite}
\usepackage{amsmath,amssymb,amsfonts}
\usepackage{graphicx}
\usepackage{textcomp}
\usepackage{xcolor}
\usepackage{float}
\usepackage{balance}
\usepackage[shortcuts,acronym]{glossaries}
\usepackage{booktabs}
\usepackage[bookmarks=false]{hyperref}
\usepackage{stfloats}
\usepackage{subcaption}

\def\BibTeX{{\rm B\kern-.05em{\sc i\kern-.025em b}\kern-.08em
    T\kern-.1667em\lower.7ex\hbox{E}\kern-.125emX}}
           
\begin{document}

\title{
Novel Deep Neural OFDM Receiver Architectures for LLR Estimation
\vspace{-10pt}
}




\author{ \IEEEauthorblockN{ Erhan Karakoca\IEEEauthorrefmark{1}, Hüseyin Çevik\IEEEauthorrefmark{1}\IEEEauthorrefmark{2}, İbrahim Hökelek\IEEEauthorrefmark{1}\IEEEauthorrefmark{3}, Ali Görçin\IEEEauthorrefmark{1}\IEEEauthorrefmark{3}}


\IEEEauthorblockA{\IEEEauthorrefmark{1} \href{https://hisar.bilgem.tubitak.gov.tr/en/} {Communications and Signal Processing Research (HİSAR) Lab., T{\"{U}}B{\.{I}}TAK B{\.{I}}LGEM, Kocaeli, Turkey}}

\IEEEauthorblockA{\IEEEauthorrefmark{2} Department of Electronics and Communication Engineering, Yildiz  Technical University, {\.{I}}stanbul, Turkey} 

\IEEEauthorblockA{\IEEEauthorrefmark{3} Department of Electronics and Communication Engineering, Istanbul Technical University, {\.{I}}stanbul, Turkey}

Emails:
\{erhan.karakoca,
huseyin.cevik, ibrahim.hokelek\}@tubitak.gov.tr, aligorcin@itu.edu.tr 

\vspace{-15pt} 
}
\maketitle
\begin{abstract}
Neural receivers have recently become a popular topic, where the received signals can be directly decoded by data-driven mechanisms such as machine learning and deep learning. In this paper, we propose two novel neural network–based orthogonal frequency division multiplexing (OFDM) receivers performing channel estimation and equalization tasks and directly predicting log‑likelihood ratios (LLRs) from the received in‑phase and quadrature‑phase (IQ) signals. The first network, the Dual Attention Transformer (DAT), employs a state‑of‑the‑art (SOTA) transformer architecture with an attention mechanism. The second network, the Residual Dual Non‑Local Attention Network (RDNLA), utilizes a parallel residual architecture with a non‑local attention block. The bit-error rate (BER) and block-error rate (BLER) performance of various SOTA neural receiver architectures is compared with our proposed methods across different signal-to-noise ratio (SNR) levels.
The simulation results show that DAT and RDNLA outperform both traditional communication systems and existing neural receiver models. 

\end{abstract}
\begin{IEEEkeywords}
Neural Receivers, LLR, Transformers
\end{IEEEkeywords}

\section{Introduction}
Data-driven techniques such as machine learning (ML) and deep learning (DL) have seen significant advances recently, not only in foundational algorithms but also in hardware design perspective, mainly driven by the “artificial intelligence (AI) at the edge” concept. For example, tensor accelerators along with new AI algorithms make computationally challenging tasks feasible to be deployed in real-world applications. These advancements facilitate an increasing number of AI deployments at the edge~\cite{peltonen20206g}.
The wireless communication community has also been heavily investing in AI/ML-based applications, where they can either build new foundational models or adapt existing ones from other domains by training them using wireless communication datasets. 

Considering advanced radio access network (RAN) technologies along with the usage of larger bandwidths in the upcoming 3GPP releases, signal processing requirements of enormous data in real-time make a traditional mathematical-based system design with cascaded communication blocks consisting of heuristic algorithms prohibitively expensive. This brings an AI-native system design forefront, where the AI/ML approaches can focus on the most important features of the underlying system by appropriately setting the weights of the trained models which provide a direct mapping between transmitted (input) and received (output) information.  In addition, the AI/ML models are suitable for revealing highly complex features for several advanced RAN functions, where hardware accelerators play a critical role in processing enormous data in parallel~\cite{samad2020white}.
As long as high-quality data is available, the model is trained and deployed at the edge. The 3GPP has already started to study AI/ML adaptation for wireless systems~\cite{lin2024overview}.

Neural receivers have been received significant attention recently, where signals are decoded by performing channel estimation, equalization, and symbol de-mapping tasks of the conventional receivers by the AI model.
For example, in~\cite{Honkala2021}, a fully convolutional deep neural network, DeepRx, is introduced to execute the whole receiver pipeline from frequency domain signal processing to channel decoding. Similarly, in~\cite{Pihlajasalo2021}, HybridDeepRx proposes residual deep learning modules including both time and frequency domain operations for demodulating orthogonal frequency division multiplexing (OFDM) signals under challenging conditions such as a high level of non-linear distortion.
Furthermore, it is shown in~\cite{ait2021end} that an end-to-end neural receiver for OFDM systems under realistic fading channels can significantly reduce pilot overhead by providing a similar bit-error rate (BER) performance compared to conventional pilot‑based systems.

The recent studies on a tensor native wireless communication~\cite{hoydis2022sionna} and Nvidia's ray tracing based simulation framework Sionna~\cite{hoydis2023sionna} provide important advancements in the context of neural receivers. The potential of neural receiver architectures has been demonstrated in~\cite{cammerer2023neural} and~\cite{Wiesmayr2024}, where neural receivers are utilized in 5G multi-user multiple input-multiple output (MIMO) scenarios in real-time while adhering to the standard compliance. Deep learning-based channel estimation methods~\cite{Xu2024}, incorporating belief knowledge and transfer learning strategies~\cite{uyoata2024transfer} for single input-multiple output (SIMO) receivers~\cite{Luostari2024} further improve performance by adapting to realistic channel conditions, as verified by aerial experiments~\cite{Luostari2024}. In addition, interpretability analysis~\cite{10888682} and performance verification frameworks for AI-specific transceiver actions~\cite{Soltani2025} provide important guidance for the robustness, safety, and operational reliability of these models.
Also,~\cite{saleem2024transrx} showed neural receiver performance can be further improved by adopting new DL models such as transformers.

Motivated by these, we propose two novel neural receiver architectures developed in Python using the Sionna framework that can be adapted to standard-based wireless communication systems. The main contributions of the paper can be summarized as:  
\begin{figure*}[!t]   
  \centering
  \includegraphics[width=1\linewidth]{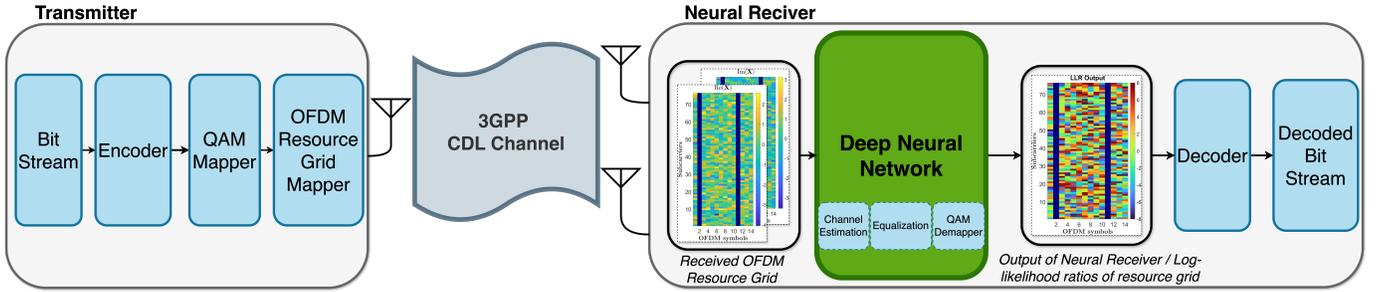}
  \caption{System Model of conventional and neural network based receiver.}
  \vspace{-15pt}
  \label{fig:system_model}
\end{figure*}
\begin{itemize}
\item We introduce two deep learning models for neural receivers that significantly improve BER and block-error rate (BLER) performance compared to the conventional Transformers. Proposed models also demonstrate notable improvements over other state-of-the-art (SOTA) models. Additionally, we provided the computation times of each model with the conventional least square (LS) estimation receiver.
\item We then explore the architecture of these models, carefully going over each element of their construction. We highlight how each element contributes to feature extraction and learning at the global and local levels that eventually affect BER and BLER.
\item Lastly, we will make the source code of our proposed models together with other SOTA model implementations publicly available\footnote{\url{https://github.com/Hisar-Research/sionna-nn-receiver-playground}}. This allows researchers and practitioners to explore, test, and expand upon our findings.
\end{itemize}
\textbf{\textit{Notation:}} 
Throughout this paper, vectors are denoted by bold lowercase letters (e.g., $\mathbf{x}$), and matrices or tensors by bold uppercase letters (e.g., $\mathbf{X}$). The complex Gaussian distribution with mean $\mu$ and variance $\sigma^2$ is denoted by $\mathcal{CN}(\mu, \sigma^2)$. $\mathbf{I}$ is the identity matrix, $\odot$ denotes the Hadamard (element-wise) product, and $*$ denotes convolution. Transpose and Hermitian transpose are represented by $^{\mathrm{T}}$ and $^{\mathrm{H}}$, respectively. Flattened tensors are indicated with an overline (e.g., $\overline{\mathbf{X}}$). Attention scores are denoted as $\mathbf{A}_{(\cdot)}$, and extracted features as $\mathbf{F}_{(\cdot)}$ of a given layer as a subindex.

\section{System Model and Background}

We consider a SIMO OFDM uplink communication system with $N_{BS}$ antennas at the Base Station (BS) as depicted in Fig.~\ref{fig:system_model}. A 3GPP Clustered Delay Line (CDL) channel model is applied in the frequency domain.
Rather than just expressing received OFDM subcarriers, given that the transmitted resource grid (RG) is $\mathbf{X}_{\text{OFDM}} \in \mathbb{C}^{N_{\text {sym}} \times N_{\text {sc}}}$ where $N_{\text{sym}}$ is the number of symbols and $N_{\text{sc}}$ is the number of subcarriers in the OFDM system, we denote received RGs at the receiver of the BS as 
\begin{equation}
\mathbf{Y}_{\text{OFDM}}=\mathbf{H} \odot \mathbf{X}_{\text{OFDM}}+\mathbf{N},
\label{eq:received_resource_grid}
\end{equation}
where $\mathbf{Y}_{\text{OFDM}} \in \mathbb{C}^{N_{\text {sym}} \times N_{\text {sc}} \times N_{\text{BS}}}$ is the received RG at the BS, containing received symbols across subcarriers, OFDM symbols, and receive antennas. $\mathbf{H} \in \mathbb{C}^{N_{\text {sym }} \times N_{\text {sc }} \times N_{\text {BS}}}$ is the uplink SIMO channel in the frequency domain. Each element $\mathbf{H}_{k, s}^{(n)}$ represents the channel gain for the $k$-th symbol and $s$-th OFDM subcarrier at the $n$-th BS antenna. Finally, $\mathbf{N} \sim \mathcal{C N}\left(\mu, \sigma^2 \mathbf{I}\right)$ is the AWGN noise at the BS.


From that point, after receiving OFDM RG, the goal of this study is to estimate the LLR values of received constellation points with DL techniques. However, before proceeding with the DL model, we will give traditional receiver blocks required to estimate LLR values. 
The UE transmits pilot symbols at specific subcarriers and time slots, forming a pilot pattern in the OFDM RG $\mathbf{X}_p$. The received pilot signals at the BS are
\begin{equation}
\mathbf{Y}_p=\mathbf{H}_p \odot \mathbf{X}_p+\mathbf{N}_p,
\label{eq:pilot_rb_estimate}
\end{equation}
and the LS channel estimate at the pilot locations is given by,
\begin{equation}
\hat{\mathbf{H}}_p=\frac{\mathbf{Y}_p}{\mathbf{X}_p}
\end{equation}
Since pilots are available only at specific time-frequency positions, the full channel matrix $\hat{\mathbf{H}}$ is obtained by interpolation over the entire OFDM RG using linear interpolation. Following channel estimation, the BS performs linear minimum mean square error (LMMSE) equalization to mitigate fading and noise. The LMMSE equalization matrix for each symbol $k$ and subcarrier $s$ is
\begin{equation}
\mathbf{W}_{k,s}^{\mathrm{LMMSE}}=\left(\hat{\mathbf{H}}_{k, s}^\mathrm{H} \hat{\mathbf{H}}_{k, s}+\frac{\sigma^2}{E_s} I\right)^{-1} \hat{\mathbf{H}}_{k, s}^\mathrm{H},
\end{equation}
and the equalized RG can be written as
\begin{equation}
\hat{\mathbf{X}}_{\text{OFDM}}=\mathbf{W}^{\mathrm{LMMSE}} \mathbf{Y}_{\text{OFDM}}
\end{equation}
where  $\mathbf{W}^{\mathrm{LMMSE}} \in \mathbb{C}^{N_{\text {sym }} \times N_{\text {sc }} \times N_{\mathrm{BS}}}$ is the equalization weight matrix for the entire RG and $\hat{\mathbf{X}}_{\text{OFDM}} \in \mathbb{C}^{N_{\text {sym }} \times N_{\text {sc}}}$ is the equalized symbol grid. After equalization, the equalized OFDM symbols are converted into bit-level LLRs. The LLR for the $i$-th bit considering a quadrature amplitude modulation (QAM) symbol is computed as:
\begin{equation}
\mathbf{L}_i(k, s)=\log \frac{P\left(b_i=1 \mid \hat{\mathbf{X}}^{k, s}_{\text{OFDM}}\right)}{P\left(b_i=0 \mid \hat{\mathbf{{X}}}^{k, s}_{\text{OFDM}}\right)}
\label{eq:llr_estimate}
\end{equation}
where $\mathbf{L} \in \mathbb{R}^{N_{\text {sym }} \times N_{\text {sc }} \times N_{\text {bits }}}$ is the LLR grid, $P(b_i=1 \mid \hat{\mathbf{X}}_{\text{OFDM}})$ and $P(b_i=0 \mid \hat{\mathbf{X}}_{\text{OFDM}})$ are conditional probability functions for each possible bit values. Finally, the LLRs are fed to the low-density parity-check (LDPC) decoder, which decodes data and reconstructs the transmitted bits. 

\begin{figure*}[ht!]
    \centering
    \includegraphics[width=1\linewidth]{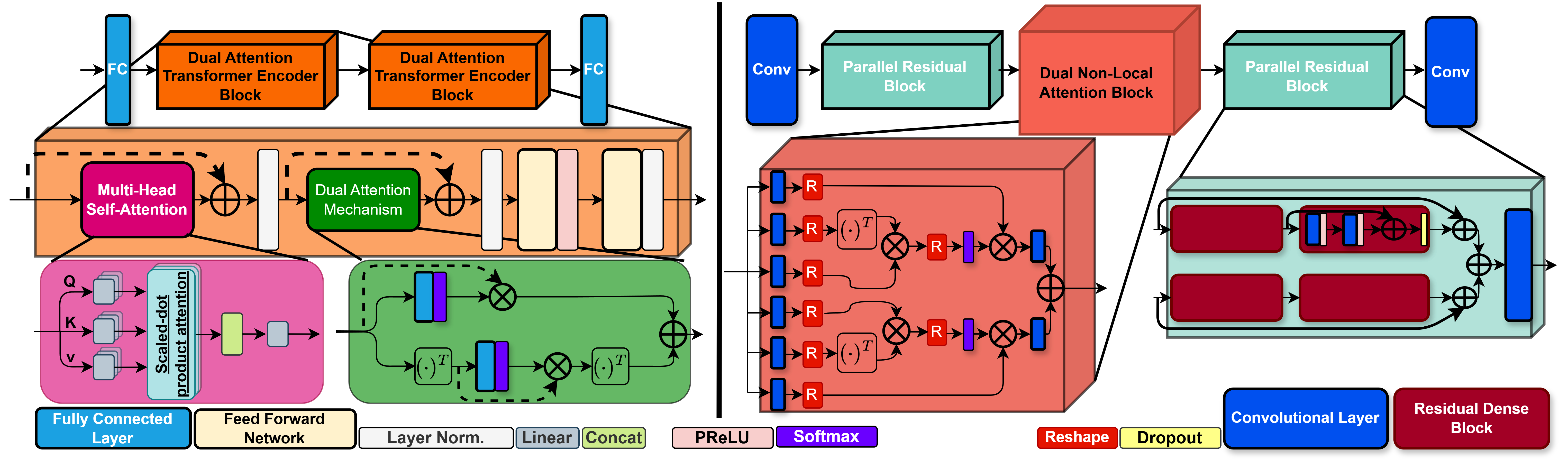}
    \caption{The DAT (left) and RDNLA (right) LLR estimation architectures for neural receiver.}
    \label{fig:enter-label}
    \vspace{-15pt}
\end{figure*}

\section{Proposed DL-Based LLR Estimation Methods}
In this section, we introduce two neural network architectures designed to directly estimate soft LLR values from the received OFDM RG $\mathbf{Y}$. The goal is to train a deep learning model that enhances BER-BLER performance by learning an effective mapping from the network input $\mathbf{Y_{\text{in}}}$ to the corresponding soft LLRs $\hat{\mathbf{L}}$. The details of this mapping function will be presented in the subsequent sections.

\subsection{Input Preprocessing and Embedding}
The received complex-valued OFDM RG
$
\mathbf{Y}_{\text{OFDM}} \in \mathbb{C}^{N_{\text{sub}}\times N_{\text{sym}}\times N_{\mathrm{BS}}},
$
is separated into real and imaginary components, producing a real-valued tensor, 
$
\mathbf{Y}^{real}_{\text{OFDM}},\;\mathbf{Y}^{imag}_{\text{OFDM}} \in \mathbb{R}^{N_{\text{sub}}\times N_{\text{sym}}\times N_{\mathrm{BS}}}.
$
In addition to the received signal samples, noise information in dB is appended as an auxiliary input, enabling the network to leverage the signal-to-noise ratio (SNR) context when producing LLRs. The noise information is broadcasted across the $(N_{\text{sym}}, N_{\text{sc}}, N_{\text{BS}})$ grid so that each network input token sees the same noise level value. Hence, the overall real-valued input $\mathbf{Y_{in}}$ can be denoted as
\begin{equation}
    \mathbf{Y_{in}} = [\mathbf{Y}_{\text{OFDM}}^{real}, \mathbf{Y}_{\text{OFDM}}^{imag}, \operatorname{log}_{10}(\boldsymbol{\mathrm{N_0}})], 
    \label{eq:y_in}
\end{equation}
and has a shape of $\in \mathbb{R}^{N_{\text{sym}} \times N_{\text{sc}} \times(2N_{\text{BS}} + 1)}$. 
\subsection{Dual Attention Transformer}

The first architecture is a Dual Attention Transformer (DAT) designed to exploit both spatial (time-frequency) and channel dependencies (correlation between receive antennas) in the OFDM RG. Compared to a standard transformer encoder, DAT incorporates two parallel attention paths, one focusing on the spatial dimension of tokens and another focusing on the channel dimension to improve LLR estimation accuracy.
Firstly, flattening the time–frequency axes of $\mathbf{Y_{in}}$ results,
\begin{equation}
\overline{\mathbf{Y}}_{\text{in}} \in \mathbb{R}^{N_T \times N_F}, \ N_{T}=N_{\text{sym}} N_{\text{sc}}, \ N_{F}=2 N_{\text{BS}}+1. 
\end{equation}
A learned fully connected linear layer embeds these features into a D-dimensional model space, 
\begin{equation}
    \mathbf{Z}=\text{FC}(\overline{\mathbf{Y}}_{\text{in}}) =  \overline{\mathbf{Y}}_{\text{in}} \mathbf{W}+\mathbf{b} \quad \in \mathbb{R}^{N_T \times N_D}
\end{equation}
where $\mathbf{W}$ denotes weights and $\mathbf{b}$ denotes bias and these are learnable parameters. Each block comprises Multi-Head Self-Attention (MHSA) and a Dual Attention Mechanism (DAM). 
Once tokenized, the data is fed through multiple Dual Attention Transformer Encoder (DAE) blocks. 
\subsubsection{Multi-Head Self Attention} First of all, MHSA captures pairwise correlations among tokens, producing attention-weighted features from $\mathbf{Z}$,
\begin{equation}
    \mathbf{Z}_{(l)} = \operatorname{MHSA}(\mathbf{Z}_{(l-1)}).
\end{equation}
For an input sequence $\mathbf{Z}$, the self-attention mechanism projects $\mathbf{Z}$ using a learnable projection matrix $\mathbf{U}_{q k v} \in \mathbb{R}^{N_T \times 3N_D}$ into
\begin{equation}
\begin{aligned}
\mathbf{U}_{\mathrm{QKV}}
&= \bigl[\mathbf{W}^Q,\;\mathbf{W}^K,\;\mathbf{W}^V\bigr],\ \\
[\mathbf{Q},\,\mathbf{K},\,\mathbf{V}]
&= \mathbf{Z}\,\mathbf{U}_{\mathrm{QKV}}
\;\in\;\mathbb{R}^{N_T\times 3N_D}.
\end{aligned}
\end{equation}
The resulting matrix separated into query $\mathbf{Q}$, key $\mathbf{K}$, and value $\mathbf{V}$ matrices where they are used to calculate attention scores by taking the scaled-dot product between the queries and keys. The softmax function is then applied row-wise to normalize the attention scores, ensuring that each row forms a probability distribution over the input tokens, and the result is multiplied by the $\mathbf{V}$~\cite{vaswani2017attention}  which are expressed as 
\begin{equation}
\begin{aligned}
\operatorname{MHSA} & =[\operatorname{head}_1, \ldots, \operatorname{head}_h] \mathbf{W}^O,\\
\text { where $\operatorname{head}_i$ } & =\underbrace{\operatorname{softmax}\left(\mathbf{Q}_i \mathbf{K}_i^T / \sqrt{d_k}\right) \mathbf{V}_i}_{\text{scaled-dot product attention}},
\label{eq:scaled-dot-product-attention}
\end{aligned}
\end{equation}
where $\mathbf{W}^O$ are trainable parameters of the heads. Finally, a residual connection and layer‐norm applied before the DAM,
\begin{equation}
\mathbf{U}=\operatorname{LN}(\mathbf{Z}_{l}+\mathbf{Z
}_{(l-1)}) \in \mathbb{R}^{N_T \times N_D}.
\end{equation}
\subsubsection{Dual-Attention Mechanism}
DAM applies first a dense layer with number of D layers, and two parallel spatial (token-wise) and channel (feature-wise) attention operations. The computation of channel-wise attention given as
\begin{equation}
\mathbf{A}_{c} = \operatorname{softmax}(\mathbf{W}_{c}\mathbf{U}^\mathrm{T} + \mathbf{b}_{c}) \in \mathbb{R}^{N_D\times N_D},
\end{equation}
\begin{equation}
    \mathbf{V}_c=\mathbf{A}_c \mathbf{U}^\mathrm{T}, \quad \mathbf{C}=\mathbf{V}_c^\mathrm{T} \in \mathbb{R}^{N_T \times N_D}.
\end{equation}
The computation of spatial-wise attention with T-units dense layer is given as
\begin{equation}
\mathbf{A}_{s} = \operatorname{softmax}(\mathbf{U}\mathbf{W}_s + \mathbf{b}_{s}) \in \mathbb{R}^{N_T\times N_T},
\end{equation}
\begin{equation}
    \mathbf{S}=\mathbf{A}_{s} \mathbf{U}, \quad 
\end{equation}
where $\mathbf{S} \in \mathbb{R}^{N_T \times N_D}$ and a layer normalization is added to the end of the DAM and it can be represented as 

\begin{equation}
\begin{aligned}
    \operatorname{DAM}(\mathbf{U}) =& \operatorname{LN}(\mathbf{U} + \operatorname{\mathbf{A}}_{\operatorname{DAM}}) = \mathbf{V}_{\operatorname{DAM}}, \\ \text{where} \ \mathbf{A}_{\operatorname{DAM}} =& \mathbf{C}+{\mathbf{S}}.
\end{aligned}
\end{equation}
In the last part of the DAE, feed-forward networks (FFNs) are used to match the desired shape of the following DAE. These last FFNs also give flexibility to networks to plug in different shape parameters to play with networks to observe performance. Furthermore, to ensure stable gradients and to preserve both global and local level extracted features, skip connections are used in every internal block of the DAE; otherwise, the gradients would vanish due to the several attention mechanisms. Then, a $\operatorname{LN}$ is used to prevent internal covariance shift. 
\begin{equation}
    \mathbf{Z_{out}}=\operatorname{reshape(}\operatorname{FC}(\operatorname{LN}(\operatorname{FFN}(\operatorname{PReLU}(\operatorname{FFN}(\mathbf{V}_{\operatorname{DAM}}))))))
\end{equation}
Finally, the last dense layer unit size is set to $N_{\text{bits}}$ in the last FC layer to match the desired shape $\mathbf{L}$, where the output is reshaped into $\mathbf{Z_{out}} \in \mathbb{R}^{N_{\text{sym}} \times N_{\text{sc}} \times N_{\text{bits}}} $.
In short, DAT emphasizes learning token-wise (spatial) and feature-wise (channel) correlations via self-attention and dual attention blocks. It is particularly adept at capturing long-range dependencies across subcarriers and symbols.

\subsection{Residual Dual Non-Local Attention Network}
RDNLA combines the benefits of residual connections and non-local attention. Residual blocks capture local and multi-level features, while non-local blocks provide global dependency modeling. A dual strategy further splits attention into spatial-wise and channel-wise branches. 
Firstly, a convolutional layer extracts low-level features and maps the last channel into $C$ channel features as given,
\begin{equation}
    \mathbf{F_0}  = \operatorname{PReLU}\bigl(\mathbf{W_0}*\mathbf{Y_{in}} + \mathbf{b}_0\bigr)\ \in \mathbb{R}^{N_{\text{sym}} \times N_{\text{sc}} \times C},
    \label{eq:convolution}
    \end{equation}
where $\mathbf{Y_{in}}$ the input, as given in Eq.~\eqref{eq:y_in}, $\mathbf{F_0}$ is the output of the layer. Also, $C$ is the growth rate parameter for the network which can be tuned. For easy interoperability, let's denote this operation in Eq.~\eqref{eq:convolution} as $\operatorname{Conv_1}$ where the subindex is the number of convolutional layers applied.

\subsubsection{Parallel Residual Block}
The network then applies a Parallel Residual Block (PRB), which merges features from two parallel arms, each containing Residual Dense Blocks (RDBs). Each RDB is defined as
\begin{equation}
 \mathbf{F}_{\text{RDB}}= \mathrm{Dropout}\bigl(\operatorname{\mathbf{F_0}+Conv_2}(\mathbf{F_0})\bigr).   
\end{equation}
Each arm in the PRB is composed of two sequential RDBs with skip connections around them. The outputs of the two arms are summed and then fused via a convolution, given as 
\begin{equation}
    \mathbf{A}_i = \operatorname{RDB}^{(2)}(\operatorname{RDB}^{(1)}(\mathbf{F_0}))+ \mathbf{F_0}, \quad i=1,2,
\end{equation}
\begin{equation}
     \mathbf{F}_{\text{PRB}} = \operatorname{Conv}_1\bigl(\mathbf{A}_1 + \mathbf{A}_2\bigr) \in \mathbb{R}^{N_{\text{sym}}\times N_{\text{sc}}\times C}.
\end{equation}

\subsubsection{Dual Non‑Local Attention Block}

Given an input feature tensor \(\mathbf{X}\in\mathbb{R}^{N_{\text{sym}}\times N_{\text{sc}}\times C}\), Dual Non‑Local Attention Block (DNLA) first applies three parallel \(1\times 1\) convolutions to produce the lower-dimensional embeddings, 
\begin{equation}
\mathbf{\Theta}= \operatorname{Conv_1}(\mathbf{X}),\;\mathbf{\Phi}=\operatorname{Conv_1}(\mathbf{X}),\;\mathbf{G}=\operatorname{Conv_1}(\mathbf{X}),
\end{equation}
where their shape is $\mathbb{R}^{N_{\text{sym}}\times N_{\text{sc}}\times C'}$ with \(C'=C/2\). By flattening the spatial dimensions into \(N_T={N_{\text{sym}}\cdot N_{\text{sc}}}\) tokens, each of these feature maps is reshaped into a matrix of size \( \mathbb{R}^{N_T\times C'}\).
In the spatial attention branch, we compute the attention scores as given, 
\begin{equation}
\mathbf{A}_s = \mathrm{softmax}\!\bigl(\overline{\mathbf{\Theta}}\overline{\mathbf{\Phi}}^{\mathrm{T}}\bigr)\;\in\;\mathbb{R}^{N_T\times N_T},
\end{equation}
where \(\overline{\mathbf{\Theta}},\overline{\mathbf{\Phi}}\) is the flattened versions of embeddings and information is aggregated from all spatial locations,
\begin{equation}
\mathbf{Z}_s=\mathbf{A}_s \overline{\mathbf{G}} \in \mathbb{R}^{N_T \times C^{\prime}}
\end{equation} 
yielding a token matrix that is then reshaped back to \(N_{\text{sym}}\times N_{\text{sc}}\times C'\) and passed through a final \(1\times1\) convolution to restore the channel dimension to \(C\).
\begin{equation}
\mathcal{S}(\mathbf{X})=\operatorname{Conv_1}(\overline{\mathbf{Z}}_s)\in \mathbb{R}^{N_{\text{sym}} \times N_{\text{sc}} \times C} .
\end{equation}
Concurrently, we form the channel-wise attention scores as,
\begin{equation}
   \mathbf{A}_c = \mathrm{softmax}\!\bigl(\overline{\mathbf{\Theta}}^{\mathrm{T}}\,\overline{\mathbf{\Phi}}\bigr)\;\in\;\mathbb{R}^{C'\times C'}, 
\end{equation}
which are then aggregated to capture global channel correlations via 
\begin{equation}
\mathbf{Z}_c=\mathbf{A}_c \overline{\mathbf{G}} \in \mathbb{R}^{ C^{\prime} \times N_T}.
\end{equation}
Later, the resulting output is projected back by 
\begin{equation}
\mathcal{C}(\mathbf{X})=\operatorname{Conv_1}(\operatorname{reshape}(\mathbf{Z}_{c}^{\mathrm{T}}))\in \mathbb{R}^{N_{\text{sym}} \times N_{\text{sc}} \times C} .
\end{equation}
Lastly, the spatial and channel attention paths are summed and added to the original input via a residual connection,  
\begin{equation}
\mathbf{F}_{\mathrm{DNLA}}
= \mathbf{X}
\;+\;\mathcal{S}(\mathbf{X})
\;+\;\mathcal{C}(\mathbf{X}), \; \in \mathbb{R}^{N_{\text{sym}} \times N_{\text{sc}}\times C},
\end{equation}
where \(\mathcal{S}(\mathbf{X})\) and \(\mathcal{C}(\mathbf{X})\) denote the restored-channel outputs of the spatial and channel branches, respectively.  This dual-path design allows each location to attend to all others across both spatial and channel dimensions, while the skip connection preserves identity information.  
The network’s output can be obtained by adding the output of the first convolutional layer to the output of each block and then passing through the last convolutional layer with $N_{\text{bits}}$ filters, as shown below
\begin{equation}
\mathbf{Z}_{\text {out }}=\operatorname{Conv_1}\left(\mathbf{F}_{\mathrm{PRB}_2}+\mathbf{F}_0\right) \ \in \mathbb{R}^{N_{\text{sym}} \times N_{\text{sc}} \times N_{\text{bits}}} .
\end{equation}
In summary, DAT and RDNLA predict the soft-LLR values for given batch examples, 
\begin{equation}
    \hat{\mathbf{L}}=f_\theta(\mathbf{Y_{in}}) \in\mathbb{R}^{N_{\text{batch}} \times N_{\text{sym}} \times N_{\text{sc}} \times N_{\text{bits}}},
    \label{eq:mapping_soft_llr}
\end{equation}
and each element of $\hat{\mathbf{L}}$ corresponds to the LLR of the $i$-th coded bit at symbol $k$, subcarrier $s$, and batch sample $m$.

\subsection{Model Training}
The given models are trained using the bit-metric decoding (BMD) rate, which serves as an information theoretic objective for bit-interleaved coded modulation (BCIM) systems. The BMD rate is computed from the transmitted coded bits and the predicted LLRs, and its optimization ensures accurate soft-bit estimation for subsequent decoding. 
The binary-cross entropy (BCE) loss function compares the predicted LLRs with the true transmitted coded bits over the entire OFDM 
RG, which is given by

\begin{equation}
\mathcal{L}_{\mathrm{BCE}}(B, \hat{L})=\mathbb{E}[B \log \sigma(\hat{L})+(1-B) \log (1-\sigma(\hat{L}))],
\label{eq:batch_BCE_loss_training}
\end{equation}
where $B$ represents the transmitted ground truth bits and $\hat{L}$ represents the predicted LLRs, finally $\sigma(\hat{L})$ is the sigmoid activation ensuring LLR interpretation as probabilities. We used the expectation operator $\mathbb{E}[\cdot]$ in loss function for conciseness, which expresses averaging over all dimensions. We used the Adam optimizer for gradients. The goal of the model is to learn the optimal parameters that minimize the loss function in Eq. \eqref{eq:batch_BCE_loss_training}, formulated as
\begin{equation}
\theta^*=\arg \min _\theta \mathcal{L}_{\mathrm{BCE}}\left(B, f_\theta\left(\mathbf{Y_{in}}\right)\right),
\end{equation}
where $\theta$ represents the learnable parameters of the model and $f_\theta(\cdot)$ is the learned mapping function, also known as the soft LLR estimator. 

\begin{figure*}[t]
\centering
\subfloat[]{\label{fig:ber}
\includegraphics[width=0.5\linewidth]{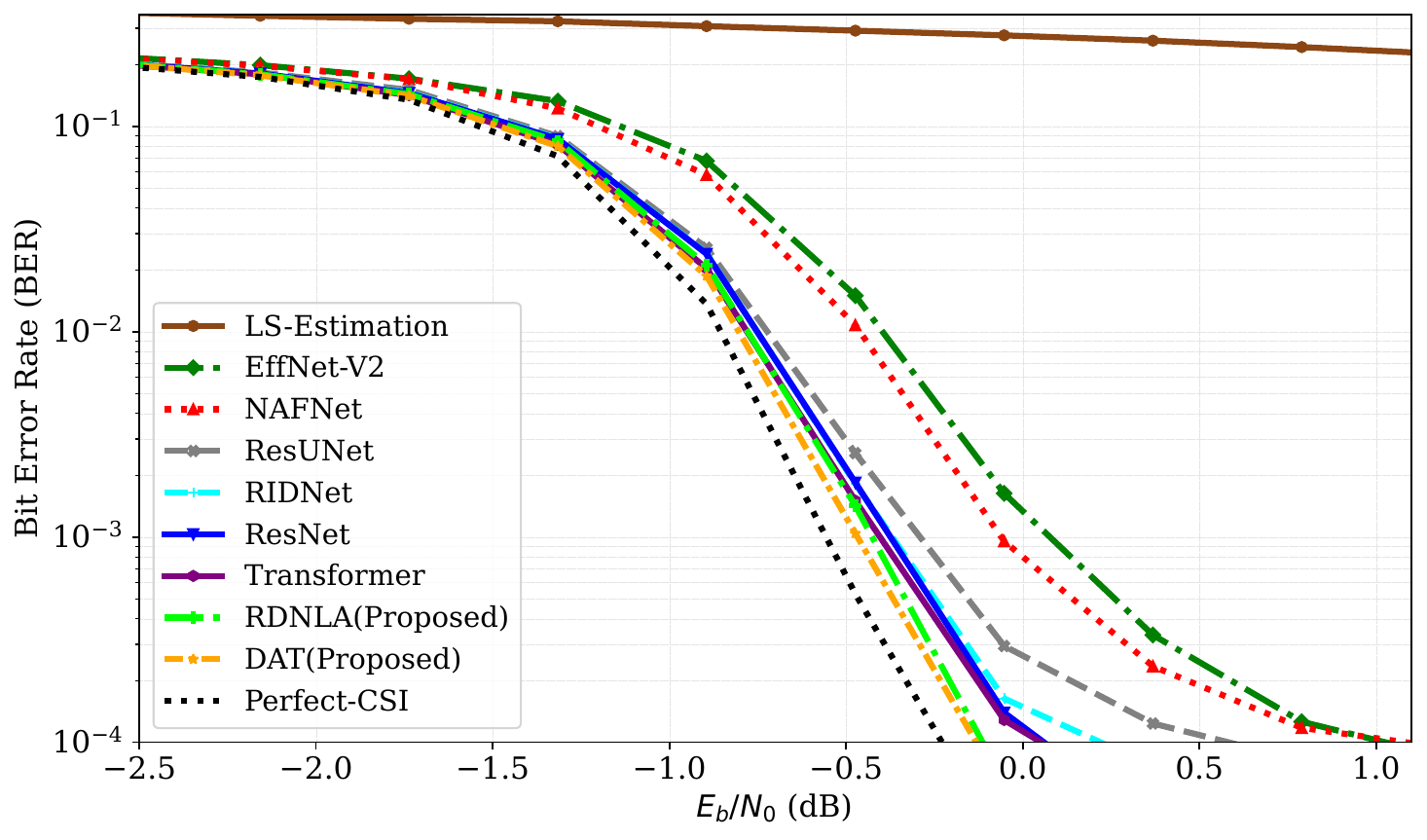}}
\subfloat[]{\label{fig:bler}
\includegraphics[width=0.5\linewidth]{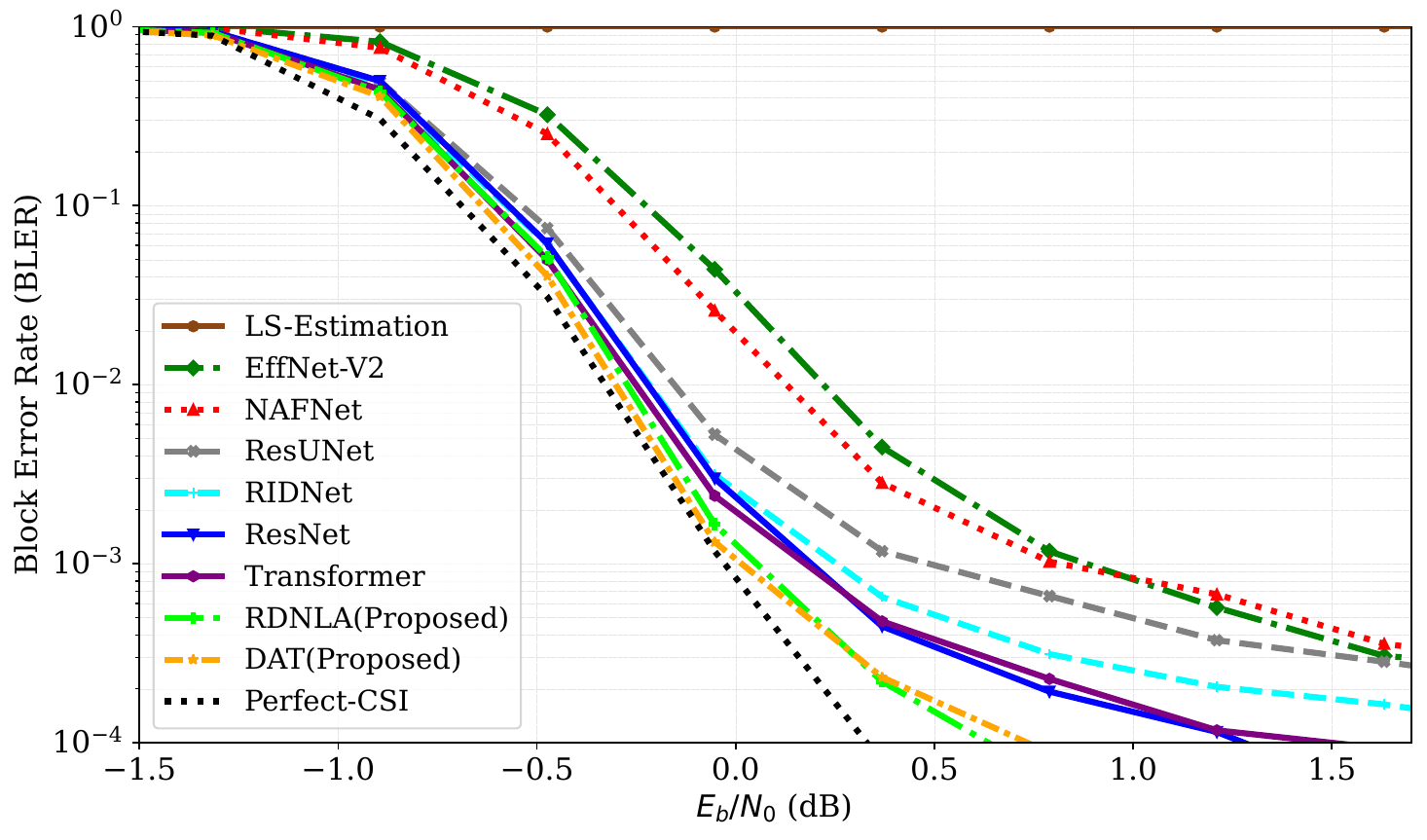}}
\vspace{-1pt}
\caption{BER (a) and BLER (b) of the proposed neural receivers with SOTA models at varying noise levels.
}
\label{fig:systemmodel2}
\vspace{-15pt}
\end{figure*}

\section{Simulation Experiments and Results}
This section presents the simulation experiments and results using Nvidia's open-source Sionna library. A wireless communication system consisting of a single antenna UE ($N_{\mathrm{UE}}=1$)  and a BS with two antennas $N_{\mathrm{BS}}=2$ in Fig.~\ref{fig:system_model} is generated using a single stream configuration. The OFDM system consists of 76 subcarriers and 14 OFDM symbols per frame, where 2 symbols are allocated for pilots. A cyclic prefix of 6 samples is used to mitigate inter-symbol interference. The signal is modulated using QAM, while LDPC encoding and decoding are applied at a coding rate of $R_C=0.5$. The proposed two neural receiver architectures, namely DAT and RDNLA, are implemented in addition to eight existing SOTA DL architectures from the literature. 
BER and BLER for all ten architectures are plotted as functions of $E_b / N_0$; for clarity, the BER curves span from -2.5 dB to 1.5 dB, while the BLER curves cover -1.5 dB to 2 dB.
Each performance result corresponds to the average of $3200$ Monte-Carlo experiments, with the target block error run set to 1000. 
For training, the batch size is set to 128 while the number of epochs is $10^5$.



Fig.~\ref{fig:ber} shows the BER results of ten different receiver models including our proposed DAT and RDNLA models under different $E_b/N_0$ values. 
Two separate performance clusters of receiver architectures emerge in the figure. The first group (ResNet~\cite{ait2021end}, Transformer~\cite{saleem2024transrx}, ResDNLA, DAT) achieves the BER results below $10^{-4}$ even $E_b/N_0$ is lower than 0.5 dB. This success is due to their architectural components such as local, and global attention blocks, and residual connections, which effectively capture complex and global relationships and enhance feature representation for neural receiver tasks. The second group (EffNet-V2, NAFNet ResUNet, RIDNet~\cite{tan2021efficientnetv2,chen2022simple,diakogiannis2020resunet,anwar2019real}) have demonstrated lower BER compared to their counterpart, especially EffNet-V2, NAFNet and ResUNet.  
Although these models perform well in semantic segmentation and incorporate skip connections and conventional attention mechanisms, they underperform in LLR estimation due to their inability to effectively capture both local and global dependencies required for this task.
The first group excels at modelling both time-frequency and spatial–channel relationships, while DAT and RDNLA further outperform standard Transformer and ResNet architectures by using finely crafted attention mechanisms that effectively capture complex feature dependencies and preserve global features. DAT and RDNLA achieve a BER of $10^{-4}$ at $E_b / N_0$ below 0 dB. In contrast, the Transformer and ResNet require $E_b / N_0$ above 0 dB to reach the same. 
While the gain is modest at low $E_b / N_0$ (below -1 dB), beyond 0 dB DAT and RDNLA steadily pull ahead, as the Transformer and ResNet fall into decline in capturing local dependencies at higher SNR.



Fig.~\ref{fig:bler} shows the BLER of the mentioned receiver models including our proposed DAT and RDNLA models under different $E_b/N_0$ values. 
The BLER results are distinguished from the BER results. The BLER values of Transformer and ResNet are struggling to converge at high SNR compared to their proposed counterparts DAT and RDNLA. This shows again Transformer and ResNet are more prone to block-level errors, due to their lack of specific attention mechanisms that result in errors on subsequent element prediction. The proposed models further exploit the complex features and can correct both global and subsequent local information. 



\begin{table}[ht]
    \centering
    \renewcommand{\arraystretch}{1.2} 
    \setlength{\tabcolsep}{6pt}      
    \caption{Mean inference times per OFDM RG}
    \label{tab:model_times}
    \begin{tabular}{lc}
        \toprule
        Model & Time (ms) \\
        \midrule
        Transformer (Encoder)                    & 11.699 \\
        NAFNet                                   & 11.891 \\
        ResNet                                   & 12.032 \\
        EffNet-V2                                & 12.270 \\
        DAT                                      & 12.366 \\
        RDNLA                                    & 12.684 \\
        ResUNet                                  & 12.910 \\
        RIDNet                                   & 13.349 \\
        LS-Estimation                            & 0.14 \\
        \bottomrule
    \end{tabular}
    \vspace{-5pt}
\end{table}
Table~\ref{tab:model_times} shows that all models process an OFDM RG within a narrow window (±1.65 ms), reflecting a balanced design in layer depth and attention complexity. The Transformer is the fastest, as it relies on dense layers and self-attention, both dominated by efficient matrix multiplications. In contrast, convolutional models (ResNet, ResUNet, RIDNet, RDNLA) use many small kernel operations that are memory-bound. 
Not surprisingly, the conventional LS-estimation receiver is faster than the neural models, underscoring the trade-off between computational efficiency and performance.

\section{Discussion and Future Work}

 This paper proposes two neural network–based OFDM receivers and compares their BER and BLER performance with eight existing state-of-art models, where the DAT employs a transformer architecture with an attention mechanism while the RDNLA utilizes a parallel residual architecture with a non‑local attention block. 
 The results show that the proposed models outperform both traditional receiver systems and existing neural receivers. However, neural receivers still require efficient architectures to catch the speed of conventional receivers. In addition, the results show that BER alone is insufficient to train the model because it ignores how errors cluster, a factor that heavily impacts BLER. The models that disperse bit errors tend to achieve lower BLER even with higher BER because scattered errors are easier for the decoder to correct, whereas the clustered errors can overwhelm it and trigger entire block failures.
Thus, we will analyze the spatial and temporal patterns of decoding errors and design architectures and training schemes that promote error dispersion. In particular, we plan to incorporate error-aware loss functions and decoder-in-the-loop feedback to directly optimize BLER.

\balance
\bibliographystyle{IEEEtran}
\bibliography{main.bib}

\begin{thebibliography}{10}
\providecommand{\url}[1]{#1}
\csname url@samestyle\endcsname
\providecommand{\newblock}{\relax}
\providecommand{\bibinfo}[2]{#2}
\providecommand{\BIBentrySTDinterwordspacing}{\spaceskip=0pt\relax}
\providecommand{\BIBentryALTinterwordstretchfactor}{4}
\providecommand{\BIBentryALTinterwordspacing}{\spaceskip=\fontdimen2\font plus
\BIBentryALTinterwordstretchfactor\fontdimen3\font minus \fontdimen4\font\relax}
\providecommand{\BIBforeignlanguage}[2]{{%
\expandafter\ifx\csname l@#1\endcsname\relax
\typeout{** WARNING: IEEEtran.bst: No hyphenation pattern has been}%
\typeout{** loaded for the language `#1'. Using the pattern for}%
\typeout{** the default language instead.}%
\else
\language=\csname l@#1\endcsname
\fi
#2}}
\providecommand{\BIBdecl}{\relax}
\BIBdecl

\bibitem{peltonen20206g}
E.~Peltonen, M.~Bennis, M.~Capobianco, M.~Debbah, A.~Ding, F.~Gil-Casti{\~n}eira, M.~Jurmu, T.~Karvonen, M.~Kelanti, A.~Kliks \emph{et~al.}, ``6{G} white paper on edge intelligence,'' \emph{arXiv preprint arXiv:2004.14850}, 2020.

\bibitem{samad2020white}
A.~Samad, W.~Saad, R.~Nandana, C.~Kapseok, D.~Steinbach, B.~Sliwa, C.~Wietfeld, K.~Mei, S.~Hamid, H.-J. Zepernick \emph{et~al.}, \emph{White Paper on Machine Learning in 6G Wireless Communication Networks: 6G Research Visions, No. 7, 2020}.\hskip 1em plus 0.5em minus 0.4em\relax University of Oulu, 2020.

\bibitem{lin2024overview}
X.~Lin, ``An overview of {AI} in 3{GPP}'s {RAN} release 18: Enhancing next-generation connectivity?'' \emph{Global Communications}, vol. 2024, 2024.

\bibitem{Honkala2021}
M.~Honkala, D.~Korpi, and J.~M. Huttunen, ``Deep{R}x: Fully convolutional deep learning receiver,'' \emph{IEEE Transactions on Wireless Communications}, vol.~20, pp. 3925--3940, 6 2021.

\bibitem{Pihlajasalo2021}
J.~Pihlajasalo, D.~Korpi, M.~Honkala, J.~M. Huttunen, T.~Riihonen, J.~Talvitie, A.~Brihuega, M.~A. Uusitalo, and M.~Valkama, ``Hybrid{D}eep{R}x: Deep learning receiver for high-{EVM} signals,'' in \emph{IEEE International Symposium on Personal, Indoor and Mobile Radio Communications, PIMRC}, vol. 2021-September.\hskip 1em plus 0.5em minus 0.4em\relax Institute of Electrical and Electronics Engineers Inc., 9 2021, pp. 622--627.

\bibitem{ait2021end}
F.~Ait~Aoudia and J.~Hoydis, ``End-to-end learning for {OFDM}: From neural receivers to pilotless communication,'' \emph{IEEE Transactions on Wireless Communications}, vol.~21, no.~2, pp. 1049--1063, 2021.

\bibitem{hoydis2022sionna}
J.~Hoydis, S.~Cammerer, F.~A. Aoudia, A.~Vem, N.~Binder, G.~Marcus, and A.~Keller, ``Sionna: An open-source library for next-generation physical layer research,'' \emph{arXiv preprint arXiv:2203.11854}, 2022.

\bibitem{hoydis2023sionna}
J.~Hoydis, F.~A{\"\i}t~Aoudia, S.~Cammerer, M.~Nimier-David, N.~Binder, G.~Marcus, and A.~Keller, ``Sionna {RT}: Differentiable ray tracing for radio propagation modeling,'' in \emph{2023 IEEE Globecom Workshops (GC Wkshps)}.\hskip 1em plus 0.5em minus 0.4em\relax IEEE, 2023, pp. 317--321.

\bibitem{cammerer2023neural}
S.~Cammerer, F.~A{\"\i}t~Aoudia, J.~Hoydis, A.~Oeldemann, A.~Roessler, T.~Mayer, and A.~Keller, ``A neural receiver for {5G NR} multi-user {MIMO},'' in \emph{2023 IEEE Globecom Workshops (GC Wkshps)}.\hskip 1em plus 0.5em minus 0.4em\relax IEEE, 2023, pp. 329--334.

\bibitem{Wiesmayr2024}
\BIBentryALTinterwordspacing
R.~Wiesmayr, S.~Cammerer, F.~A. Aoudia, J.~Hoydis, J.~Zakrzewski, and A.~Keller, ``Design of a standard-compliant real-time neural receiver for {5G NR},'' 9 2024. [Online]. Available: \url{http://arxiv.org/abs/2409.02912}
\BIBentrySTDinterwordspacing

\bibitem{Xu2024}
\BIBentryALTinterwordspacing
J.~Xu, L.~Liu, X.~Wang, and L.~Chen, ``Belief information based deep channel estimation for massive {MIMO} systems,'' 6 2024. [Online]. Available: \url{http://arxiv.org/abs/2407.07744}
\BIBentrySTDinterwordspacing

\bibitem{uyoata2024transfer}
U.~E. Uyoata and R.~O. Adeogun, ``On transfer learning for a fully convolutional deep neural {SIMO} receiver,'' in \emph{2024 IEEE 100th Vehicular Technology Conference (VTC2024-Fall)}.\hskip 1em plus 0.5em minus 0.4em\relax IEEE, 2024, pp. 1--7.

\bibitem{Luostari2024}
\BIBentryALTinterwordspacing
R.~Luostari, D.~Korpi, M.~Honkala, and J.~M.~J. Huttunen, ``Adapting to reality: Over-the-air validation of {AI}-based receivers trained with simulated channels,'' 8 2024. [Online]. Available: \url{http://arxiv.org/abs/2408.04182}
\BIBentrySTDinterwordspacing

\bibitem{10888682}
M.~Tuononen, D.~Korpi, and V.~Hautamäki, ``Interpreting deep neural network-based receiver under varying signal-to-noise ratios,'' in \emph{ICASSP 2025 - 2025 IEEE International Conference on Acoustics, Speech and Signal Processing (ICASSP)}, 2025, pp. 1--5.

\bibitem{Soltani2025}
\BIBentryALTinterwordspacing
N.~Soltani, M.~Loehning, and K.~Chowdhury, ``{VERITAS}: Verifying the performance of {AI}-native transceiver actions in base-stations,'' 1 2025. [Online]. Available: \url{http://arxiv.org/abs/2501.09761}
\BIBentrySTDinterwordspacing

\bibitem{saleem2024transrx}
O.~Saleem, S.~Ribouh, M.~Alfaqawi, A.~Bensrhair, and P.~Merdrignac, ``Trans{R}x-{6G}-{V2X}: Transformer encoder-based deep neural receiver for next generation of cellular vehicular communications,'' in \emph{2024 IEEE 100th Vehicular Technology Conference (VTC2024-Fall)}.\hskip 1em plus 0.5em minus 0.4em\relax IEEE, 2024, pp. 1--7.

\bibitem{vaswani2017attention}
A.~Vaswani, N.~Shazeer, N.~Parmar, J.~Uszkoreit, L.~Jones, A.~N. Gomez, {\L}.~Kaiser, and I.~Polosukhin, ``Attention is all you need,'' \emph{Advances in neural information processing systems}, vol.~30, 2017.

\bibitem{tan2021efficientnetv2}
M.~Tan and Q.~Le, ``Efficientnetv2: Smaller models and faster training,'' in \emph{International conference on machine learning}.\hskip 1em plus 0.5em minus 0.4em\relax PMLR, 2021, pp. 10\,096--10\,106.

\bibitem{chen2022simple}
L.~Chen, X.~Chu, X.~Zhang, and J.~Sun, ``Simple baselines for image restoration,'' in \emph{European conference on computer vision}.\hskip 1em plus 0.5em minus 0.4em\relax Springer, 2022, pp. 17--33.

\bibitem{diakogiannis2020resunet}
F.~I. Diakogiannis, F.~Waldner, P.~Caccetta, and C.~Wu, ``Resunet-a: A deep learning framework for semantic segmentation of remotely sensed data,'' \emph{ISPRS Journal of Photogrammetry and Remote Sensing}, vol. 162, pp. 94--114, 2020.

\bibitem{anwar2019real}
S.~Anwar and N.~Barnes, ``Real image denoising with feature attention,'' in \emph{Proceedings of the IEEE/CVF international conference on computer vision}, 2019, pp. 3155--3164.

\end{thebibliography}


\end{document}